\documentclass[article,preprint,groupedaddress]{revtex4}
\usepackage{epsfig}
\usepackage{graphicx}
\usepackage{amssymb}

\begin{document}
\title{No-scalar-hair theorem for spherically symmetric reflecting
stars}
\author{Shahar Hod}
\affiliation{The Ruppin Academic Center, Emeq Hefer 40250, Israel}
\affiliation{ }
\affiliation{The Hadassah Institute, Jerusalem 91010, Israel}
\date{\today}

\begin{abstract}
\ \ \ It is proved that spherically symmetric compact reflecting
objects cannot support static bound-state configurations made of
scalar fields whose self-interaction potential $V(\psi^2)$ is a
monotonically increasing function of its argument. Our theorem rules
out, in particular, the existence of massive scalar hair outside the
surface of a spherically symmetric compact reflecting star.
\end{abstract}
\bigskip
\maketitle

\section{Introduction}

The no-hair theorem of Bekenstein \cite{Bek1} (see also \cite{NSO})
has revealed the intriguing fact that, in accord with Wheeler's
celebrated conjecture \cite{Whee,Car}, asymptotically flat black
holes cannot support regular self-gravitating static matter
configurations made of massive scalar fields in their external
spacetime regions \cite{NoteV,Bek20,Notensk,Hodrc,HerR}. This
important physical characteristic of black holes is often attributed
in the physics literature to the fact that the boundary of a
classical black hole (its horizon) acts as a one-way membrane that
irreversibly absorbs matter and radiation fields.

One naturally wonders whether this no-scalar-hair behavior is a {\it
unique} property of black holes? In the present paper we shall
explore the possibility of extending the no-scalar-hair theorem to
the regime of regular (that is, {\it horizonless}) curved
spacetimes. In particular, we here raise the following physically
intriguing question: Can regular compact reflecting objects (that
is, reflecting stars \cite{Noteref} which possess no event horizons)
support self-gravitating massive scalar field configurations in
their exterior spacetime regions?

In order to address this interesting question, in the present study
we shall replace the standard ingoing ({\it absorbing}) boundary
condition which characterizes the behavior of classical fields at
the horizon of a black hole \cite{Bek1,Bek20} by a reflecting ({\it
repulsive}) boundary condition at the surface of the horizonless
compact star. Our theorem, to be proved below, reveals the
intriguing fact that horizonless compact reflecting stars share the
characteristic no-scalar-hair property with asymptotically flat
black holes.

\section{The no-scalar-hair theorem for spherically symmetric compact reflecting stars}

We consider a static spherically symmetric compact reflecting object
(a reflecting star \cite{Noteref}) of radius $R$. Using the
Schwarzschild coordinates $(t,r,\theta,\phi)$, the line element of
the corresponding curved spacetime can be expressed in the form
\cite{BekMay,Noteunit}
\begin{equation}\label{Eq1}
ds^2=-e^{\nu}dt^2+e^{\lambda}dr^2+r^2(d\theta^2+\sin^2\theta
d\phi^2)\ ,
\end{equation}
where $\nu=\nu(r)$ and $\lambda=\lambda(r)$. An asymptotically flat
spacetime is characterized by the asymptotic behaviors $\nu\sim
O(r^{-1})$ and $\lambda\sim (r^{-1})$ for $r\to\infty$.

The compact star is non-linearly coupled to a real scalar field
$\psi$ with a general self-interaction potential $V=V(\psi^2)$ whose
action is given by \cite{BekMay,NoteSEH}
\begin{equation}\label{Eq2}
S=S_{EH}-{1\over2}\int\Big[\partial_{\alpha}\psi\partial^{\alpha}\psi+V(\psi^2)\Big]\sqrt{-g}d^4x\
.
\end{equation}
We shall assume a positive semidefinite self-interaction potential
for the scalar field which is a monotonically increasing function of
its argument. That is,
\begin{equation}\label{Eq3}
V(0)=0\ \ \ \text{with}\ \ \ \dot V\equiv
{{d[V(\psi^2)]}\over{d(\psi^2)}}\geq0\ .
\end{equation}
Note that the physically interesting case of a free massive scalar
field with $\dot V=\mu^2\geq0$ is covered by this form of the
self-interaction scalar potential.

From the action (\ref{Eq2}) one finds the characteristic equation
\cite{BekMay}
\begin{equation}\label{Eq4}
\partial_{\alpha}\partial^{\alpha}\psi-\dot V\psi=0\
\end{equation}
for the self-interacting scalar field. Substituting the line-element
(\ref{Eq1}) of the spherically symmetric curved spacetime into
(\ref{Eq4}), one obtains the characteristic radial differential
equation \cite{Notetag}
\begin{equation}\label{Eq5}
\psi{''}+{1\over2}\big({{4}\over{r}}+\nu{'}-\lambda{'}\big)\psi{'}-e^{\lambda}\dot
V\psi=0\
\end{equation}
for the self-interacting static scalar field.

The energy density of the self-interacting scalar field (\ref{Eq2})
is given by \cite{BekMay}
\begin{equation}\label{Eq6}
\rho=-T^{t}_{t}={1\over2}\big[e^{-\lambda}(\psi{'})^2+V(\psi^2)\big]\
.
\end{equation}
An asymptotically flat (finite mass) spacetime is characterized by
an energy density $\rho$ which approaches zero asymptotically faster
than $1/r^3$ \cite{Hodasm}:
\begin{equation}\label{Eq7}
r^3\rho(r)\to0\ \ \ \text{for}\ \ \ r\to\infty\  .
\end{equation}
Taking cognizance of Eqs. (\ref{Eq3}), (\ref{Eq6}), and (\ref{Eq7}),
one deduces that the scalar field eigenfunction is characterized by
the asymptotic boundary condition
\begin{equation}\label{Eq8}
\psi(r\to\infty)\to0\
\end{equation}
at spatial infinity. In addition, we shall assume that the scalar
field vanishes on the surface $r=R$ of the compact {\it reflecting}
star \cite{Noterefbs,PressTeu2}:
\begin{equation}\label{Eq9}
\psi(r=R)=0\  .
\end{equation}

Taking cognizance of the boundary conditions (\ref{Eq8}) and
(\ref{Eq9}), one concludes that the characteristic eigenfunction
$\psi$ of the scalar field must have (at least) one extremum point,
$r=r_{\text{peak}}$, between the surface $r=R$ of the reflecting
star and spatial infinity [that is, in the interval
$r_{\text{peak}}\in(R,\infty)]$. At this extremum point the
eigenfunction $\psi$ of the external scalar field is characterized
by the relations
\begin{equation}\label{Eq10}
\{\psi{'}=0\ \ \ \text{and}\ \ \ \psi\cdot\psi{''}<0\}\ \ \ \
\text{for}\ \ \ \ r=r_{\text{peak}}\  .
\end{equation}
Substituting (\ref{Eq3}) and (\ref{Eq10}) into the l.h.s of
(\ref{Eq5}), one finds the inequality
\begin{equation}\label{Eq11}
\psi{''}+{1\over2}\big({{4}\over{r}}+\nu{'}-\lambda{'}\big)\psi{'}-e^{\lambda}\dot
V\psi<0\ \ \ \ \text{for}\ \ \ \ r=r_{\text{peak}}\  ,
\end{equation}
in {\it contradiction} with the characteristic relation (\ref{Eq5})
of the self-interacting scalar field.

\section{Summary}

In this compact analysis, we have proved that {\it if} a spherically
symmetric compact reflecting star \cite{Noteref} can support
self-gravitating massive scalar field configurations, then the
corresponding scalar field eigenfunction $\psi$ must have an
extremum point outside the reflecting surface of the star. At this
extremum point, the scalar field eigenfunction is characterized by
the inequality (\ref{Eq11}). However, one realizes that this
inequality is in {\it contradiction} with the characteristic
identity (\ref{Eq5}) for the self-interacting scalar field. Thus,
there is no solution for the external scalar eigenfunction except
the trivial one, $\psi\equiv0$ \cite{Notefr1,Notefr2}.

We thus conclude that spherically symmetric compact reflecting
objects cannot support static bound-state configurations made of
scalar fields whose self-interaction potential $V(\psi^2)$ is a
monotonically increasing function of its argument. In particular,
our theorem rules out the existence of asymptotically flat massive
scalar hair (regular self-gravitating massive scalar field
configurations) outside the surface of a spherically symmetric
(horizonless) compact reflecting star.

Our compact theorem therefore reveals the interesting fact that {\it
horizonless} compact reflecting stars share the no-scalar-hair
property with the more familiar asymptotically flat absorbing
\cite{Notehor} black holes.

\newpage

\bigskip
\noindent
{\bf ACKNOWLEDGMENTS}
\bigskip

This research is supported by the Carmel Science Foundation. I would
like to thank Yael Oren, Arbel M. Ongo, Ayelet B. Lata, and Alona B.
Tea for helpful discussions.



\begin{thebibliography}{99}

\bibitem{Bek1} J. D. Bekenstein, Phys. Rev. Lett. {\bf 28}, 452
(1972).

\bibitem{NSO} J. E. Chase, Commun. Math. Phys. {\bf 19}, 276 (1970); C. Teitelboim,
Lett. Nuovo Cimento {\bf 3}, 326 (1972); J. D. Bekenstein, Physics
Today {\bf 33}, 24 (1980).

\bibitem{Whee} R. Ruffini and J. A. Wheeler, Phys. Today {\bf 24}, 30
(1971).

\bibitem{Car} B. Carter, in {\it Black Holes}, Proceedings of 1972 Session of Ecole d'ete de Physique Theorique,
edited by C. De Witt and B. S. De Witt (Gordon and Breach, New York,
1973).

\bibitem{NoteV} As noted in \cite{Bek20}, this interesting no-hair
property of black holes can be extended to the regime of
self-gravitating scalar fields whose self-interaction potential
$V(\psi^2)$ is a monotonically increasing function of its argument
[see Eqs. (\ref{Eq2}) and (\ref{Eq3}) below].

\bibitem{Bek20} J. D. Bekenstein, arXiv:gr-qc/9605059 .

\bibitem{Notensk} Interestingly, it has recently been
proved \cite{Hodrc,HerR} that stationary configurations made of
massive scalar fields can be supported in the external spacetime
regions of spinning black holes.

\bibitem{Hodrc} S. Hod, Phys. Rev. D {\bf 86}, 104026 (2012) [arXiv:1211.3202];
S. Hod, The Euro. Phys. Journal C {\bf 73}, 2378 (2013)
[arXiv:1311.5298]; S. Hod, Phys. Rev. D {\bf 90}, 024051 (2014)
[arXiv:1406.1179]; S. Hod, Phys. Lett. B {\bf 739}, 196 (2014)
[arXiv:1411.2609]; S. Hod, Class. and Quant. Grav. {\bf 32}, 134002
(2015) [arXiv:1607.00003]; S. Hod, Class. and Quant. Grav. {\bf 33},
114001 (2016); S. Hod, Phys. Lett. B {\bf 758}, 181 (2016)
[arXiv:1606.02306]; S. Hod and O. Hod, Phys. Rev. D {\bf 81}, 061502
Rapid communication (2010) [arXiv:0910.0734]; S. Hod, Phys. Lett. B
{\bf 708}, 320 (2012) [arXiv:1205.1872].

\bibitem{HerR} C. A. R. Herdeiro and E. Radu, Phys. Rev. Lett. {\bf 112}, 221101
(2014); C. A. R. Herdeiro and E. Radu, Phys. Rev. D {\bf 89}, 124018
(2014); C. A. R. Herdeiro and E. Radu, Int. J. Mod. Phys. D {\bf
23}, 1442014 (2014); C. L. Benone, L. C. B. Crispino, C. Herdeiro,
and E. Radu, Phys. Rev. D {\bf 90}, 104024 (2014); C. Herdeiro, E.
Radu, and H. R\'unarsson, Phys. Lett. B {\bf 739}, 302 (2014); C.
Herdeiro and E. Radu, Class. Quantum Grav. {\bf 32} 144001 (2015);
C. A. R. Herdeiro and E. Radu, Int. J. Mod. Phys. D {\bf 24},
1542014 (2015); C. A. R. Herdeiro and E. Radu, Int. J. Mod. Phys. D
{\bf 24}, 1544022 (2015); J. C. Degollado and C. A. R. Herdeiro,
Gen. Rel. Grav. {\bf 45}, 2483 (2013); P. V. P. Cunha, C. A. R.
Herdeiro, E. Radu, and H. F. R\'unarsson, Phys. Rev. Lett. {\bf
115}, 211102 (2015); C. A. R. Herdeiro, E. Radu, and H. F.
R\'unarsson, Phys. Rev. D {\bf 92}, 084059 (2015); Y. Brihaye, C.
Herdeiro, and E. Radu, Phys. Lett. B {\bf 760}, 279 (2016).

\bibitem{Noteref} We use here the term `reflecting star' to describe a physical compact object for which
the external scalar field vanishes on its surface.

\bibitem{BekMay} A. E. Mayo and J. D. Bekenstein, Phys. Rev. D {\bf 54}, 5059 (1996).

\bibitem{Noteunit} We shall use natural units in which $G=c=1$.

\bibitem{NoteSEH} Here $S_{\text{EH}}$ is the Einstein-Hilbert action.

\bibitem{Notetag} Here a prime ${'}$ denotes a derivative with respect to the radial coordinate $r$.

\bibitem{Hodasm} S. Hod, Phys. Lett. B {\bf 739}, 383 (2014) [arXiv:1412.3808].

\bibitem{Noterefbs} It is worth noting that in the vast physics literature that deals
with the famous `black-hole bomb' mechanism of Press and Teukolsky
\cite{PressTeu2}, one usually places a reflecting surface around a
black hole in order to prevent the scalar field from escaping to
infinity. On the other hand, in the present study the role of the
reflecting surface is to prevent the scalar field from entering the
central horizonless compact star.

\bibitem{PressTeu2} W. H. Press and S. A. Teukolsky, Nature {\bf
238}, 211 (1972); W. H. Press and S. A. Teukolsky, Astrophys. J.
{\bf 185}, 649 (1973).

\bibitem{Notefr1} As nicely emphasized by the anonymous referee, the result of the present paper can be 
framed in the familiar context of standard quantum mechanics. In particular, a stationary state of a standard 
one dimensional quantum mechanical problem is also characterized by the relation (\ref{Eq10}), which implies that the 
energy of the stationary quantum state is bounded from below by the potential energy at the corresponding extremum point. Note that in the terminology of quantum mechanics, our static scalar field corresponds to a zero-energy state, whereas the effective radial potential is positive [see (\ref{Eq3})].

\bibitem{Notefr2} As pointed out by the anonymous referee, it would be interesting to check whether a reflecting star can support a stationary complex scalar field configuration around it. 

\bibitem{Notehor} It is worth emphasizing again that black holes, as opposed to the
compact reflecting stars discussed in the present analysis, are
characterized by the presence of {\it absorbing} one-way membranes
(event horizons).

\end{thebibliography}
\end{document}